\documentclass[doublecol]{epl2}

\usepackage{amsmath}
\newcommand{\fat}[1]{\mathbf{#1}}
\newcommand{\fatsym}[1]{\boldsymbol{#1}}

\title{Non-equilibrium condensation and coarsening of field-driven dipolar colloids}
\shorttitle{Condensation and dynamic coarsening of field-driven dipolar colloids} 

\author{Sebastian J\"ager\inst{1} \and Heiko Schmidle\inst{1} \and Sabine H. L. Klapp\inst{1}}
\shortauthor{Sebastian J\"ager \etal}

\institute{                    
\inst{1}Institute of Theoretical Physics, Technical University Berlin,
  Hardenbergstr.~36, 10623 Berlin, Germany
}

\abstract{
In colloidal suspensions, self-organization processes can be easily fueled by
external fields. One particularly interesting class of phenomena occurs in
monolayers of dipolar particles that are driven by rotating external fields.
Here we report results from a computer simulation study of such systems
focusing on the clustering behavior also observed in recent experiments. The
key result of this paper is a novel interpretation of this pattern formation
phenomenon: We show the clustering to be a by-product of a vapor-liquid first
order phase transition. In fact, the observed dynamic coarsening process
corresponds to the spindodal demixing that occurs during such a transition.}

\begin{document}
\maketitle

\section{Introduction}
Self-assembly and self-organization processes of colloidal particles are topics
that have been receiving much attention in the recent past.

Indeed, such systems display a multitude of equilibrium and non-equilibrium
self-assembled structures, examples being lane formation \cite{dzubiella2002,
loewen2010}, shear banding \cite{kang2006}, the coiling up of magnetic chains
\cite{casic}, and the wide range of patterns observed in particles immersed in
liquid crystals \cite{ognysta2009}. Here, we are particularly interested in
colloidal systems involving dipolar interactions. Prime examples of the
resulting self-assembled structures include chain formation in constant
external fields \cite{martin4, martin6, martin3}, layer formation in rotating
fields \cite{martin3, martin6, leunissen, jaeger_klapp}, and the structure
formation of colloidal particles in triaxial fields \cite{martin5, ostermann,
Douglas2010}.

Up until recently, most works on self-organization under the influence of
time-dependent external fields focused on {\it induced} dipolar particles. A
noteworthy exception is a paper by Murashov and Patey, in which layer formation
of rotationally driven colloidal particles carrying a permanent dipole moment
was investigated \cite{murashov}.

A particularly interesting self-organization process occurs when colloidal
particles are exposed to rotating fields in a quasi-two-dimensional geometry.
In this situation, the external fields are found to induce the formation of
two-dimensional clustered structures. Not only does this work for particles in
which a dipole moment can be induced \cite{elsner, tierno2007, snoswell2006},
but also for particles carrying a permanent dipole moment. This was recently
shown by Weddemann and coworkers in an experimental study \cite{weddemann2010,
wittbracht2011}. 

In the present paper, we want to pick up on this phenomenon and provide a novel
interpretation of the two-dimensional cluster formation process. Specifically,
we will show that this self-organization process is a by-product of an
equilibrium liquid-vapor phase transition.

We investigate the colloidal system by making use of different computer
simulation techniques. In these simulations, we use dipolar particles in a
quasi-two-dimensional geometry to model the system. This means that the dipoles
can rotate freely in all the spatial directions while the translational motion
is restricted to a two-dimensional plane. We use Brownian dynamics simulations
to understand the dynamical properties of the system and
Wang-Landau-Monte-Carlo simulations to look into its phase behavior.
Additionally, to assess the influence of the solvent on the system, we take it
implicitly into account by employing Brownian dynamics simulations that include
hydrodynamic interactions.

This paper is organized as follows: After introducing the model and the
different simulation techniques, we first discuss the full non-equilibrium
``phase'' diagram indicating the region of cluster formation in the domain of
frequency and strength of the external field at constant equilibrium
thermodynamic parameters. In a next step, we investigate the influence of
hydrodynamic interactions on the formation of clusters. Then we present the
principal point of this paper: We show that the non-equilibrium cluster
formation is essentially an equilibrium phase transition. To do this, we
calculate an equilibrium phase diagram, in the construction of which an
effective non-time-dependent inter-particle interaction is used, and examine
the growth of the characteristic domain size of the clusters. The paper is then
closed with a brief summary and conclusions.

\section{Model}
To model the (dipolar) colloidal particles in our simulations we use a dipolar
soft sphere (DSS) potential, which is comprised of a repulsive part
$U^{\mathrm{rep}}$ and a point dipole-dipole interaction part  
\begin{multline}
    \label{eq:interaction}
    U^{\mathrm{DSS}}(\fat{r}_{ij}, \fatsym{\mu}_i, \fatsym{\mu}_j) =
    U^{\mathrm{rep}}(r_{ij}) \\
    - \frac{3 (\fat{r}_{ij} \cdot \boldsymbol{\mu}_i) (\fat{r}_{ij} \cdot 
    \boldsymbol{\mu}_j)}{r_{ij}^5} + \frac{\boldsymbol{\mu}_i \cdot 
    \boldsymbol{\mu}_j}{r_{ij}^3}  .
\end{multline}
In eq.~(\ref{eq:interaction}), $\fat{r}_{ij}$ is the vector between the
positions of the particles $i$ and $j$, $r_{ij}$ its absolute value, and
$\boldsymbol{\mu}_i$ is the dipole moment of the $i$th particle. The potential
$U^{\mathrm{rep}}$ is the (shifted, cf.~\cite{allentil}) soft sphere (SS)
potential for particles of diameter $\sigma$, with the unshifted SS potential
being given by $U^{\mathrm{SS}}(r) = 4 \epsilon (\sigma/r_{ij})^{12}$.

We investigate the system by making use of different simulation techniques.
First, we employ non-overdamped Brownian dynamics (BD) simulations. The
corresponding equations of motion for particles of mass $m$ and moment of
inertia $I$ are\cite{mueller2002, murashov} 
\begin{align}
    \label{eq:eom_trans}
    m \ddot{\fat{r}}_i & = \fat{F}_i^\mathrm{DSS}
    - \xi_T \dot{\fat{r}}_i + \fat{F}^\mathrm{G}_i \\
    \label{eq:eom_rot}
    I \dot{\fatsym{\omega}}_i & = \fat{T}_i^{\mathrm{DSS}}
    + \fat{T}_i^{\mathrm{ext}} - \xi_R \fatsym{\omega}_i
    + \fat{T}^{\mathrm{G}}_i ,
\end{align}
where $\xi_T$ and $\xi_R$ are the translational and rotational friction
coefficients, respectively. Consistent with earlier simulation studies of
ferrofluidic particles \cite{murashov, jaeger_klapp}, $\xi_T = 13.5 \sqrt{m
\epsilon /\sigma^2}$ and $\xi_R = 0.45 \sqrt{m \epsilon \sigma^2}$ were used.
Furthermore, $\fatsym{\omega}_i$ is the angular velocity of particle $i$,
$\fat{F}^G_i$ and $\fat{T}^G_i$ are random Gaussian forces and torques whose
variance is related to the friction coefficients \cite{murashov}, and
$\fat{T}_i^\mathrm{ext}$ are torques due to an external field. The field is
homogeneous, rotates in the dipolar monolayer, and is given by \begin{equation}
    \fat{B}(t) = B_0 ( \fat{e}_x \cos \omega_0 t + \fat{e}_y \sin \omega_0 t ),
\end{equation} where $\omega_0$ is the frequency of the field and $B_0$ its
strength.

The equations of motion (\ref{eq:eom_trans}) and (\ref{eq:eom_rot}) were
integrated with a Leapfrog algorithm \cite{allentil} using a time-step of
$\Delta t = 0.0025 (m \sigma^2 / \epsilon)^{1/2}$ and 4900 or 1225 particles,
respectively.

Further, to investigate the influence of a solvent within our implicit model,
we use a Brownian dynamics simulation that includes hydrodynamic interactions
between the particles. These interactions are incorporated up to third order in
the diffusion tensor for the translation-translation coupling, the
rotation-rotation coupling, and the translation-rotation (and vice versa)
coupling \cite{dickinson, meriguet}. The tensor describing the
translation-translation coupling is the well known Rotne-Prager tensor. The
other couplings incorporate the influence of the additional rotational degrees
of freedom. The time evolution equations that were used can be found in
\cite{meriguet} and we used 324 particles in these simulations.

Finally, to investigate the equilibrium phase behavior of the system (based on
a time-averaged potential, see eq.~(\ref{eq:dipole_avg})), we use Monte Carlo
(MC) simulations in the Grand-canonical ensemble. In general, first-order phase
transitions are plagued by a large free energy barrier separating both phases,
making unbiased sampling very inefficient. In order to overcome the barrier we
use a method proposed by Wang and Landau \cite{landau2001} which significantly
improves the sampling by adding a weight-function to the simulation. Details of
the method applied to dipolar systems in two dimensions can be found in
\cite{schmidle2011}.

In the case of BD and MC simulations we deal with the long-ranged dipolar
interactions by using the Ewald summation method\cite{Weis2003}. In our
hydrodynamic simulations, on the other hand, we only consider a single
simulation box filled with dipolar particles.

\section{Synchronization and cluster formation}
\begin{figure}[ht]
    \centering
    \includegraphics[width=83mm]{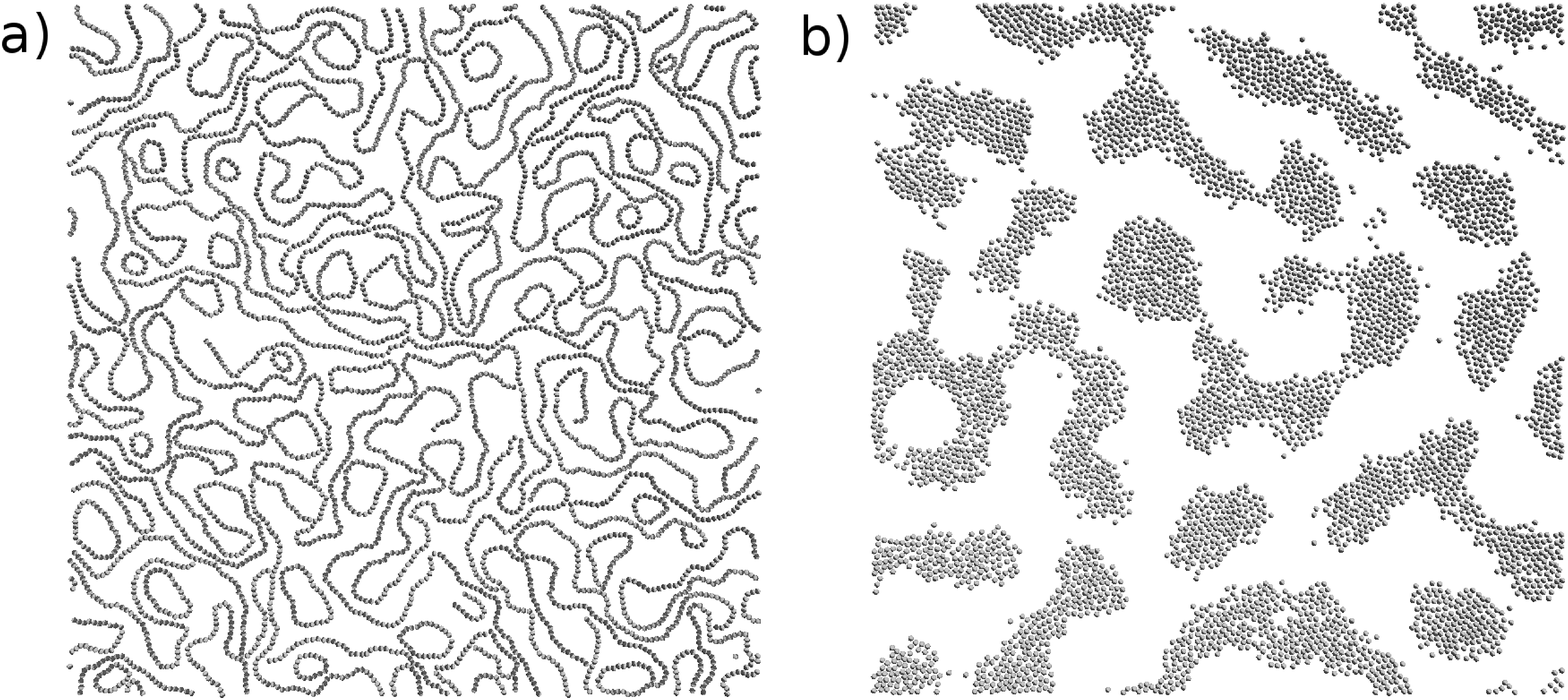}
    \caption{Snapshots showing a system in zero field (a) and exposed to a
    rotating field of strength $(\epsilon/ \sigma^3)^{1/2} B_0 = 50.0$ and
    frequency $(m \sigma^2/\epsilon)^{-1/2} \omega_0 = 20$ (b).
    Snapshot (b) was taken at time $t = 375 (m \sigma^2 / \epsilon)^{1/2}$
    after the start of the simulation. $4900$ particles were used.}
    \label{fig:snaps}
\end{figure}
A snapshot of a quasi-two-dimensional system ($\rho \sigma^2 = 0.3$, $k_B T /
\epsilon = 1.0$) of strongly coupled dipolar particles ($\lambda_{\mathrm{DD}}
= \mu^2 / k_B T \sigma^3 = 9)$ at zero field is shown in fig.~\ref{fig:snaps}a.
As is typical for such dipolar systems the particles align in a head-to-tail
configuration, which, in a two-dimensional geometry, results in the formation
of chains and rings \cite{duncan2004}.

If we expose such a system to a rotating in-plane field of sufficient strength
and frequency, we observe the particles agglomerate into two dimensional
clusters. An example of this can be seen in fig.~\ref{fig:snaps}b. The observed
clustering behavior already indicates that there are attractive interactions in
the system, which play a crucial role. The origin of these interactions can be
rationalized as follows: Averaging the dipolar inter-particle interaction
potential over one rotational period of the field under the assumptions that
the particles do not move translationally and rotate synchronously with the
field (i.e.~follow the field at constant phase difference) yields
\begin{equation}
    \label{eq:dipole_avg}
    U^{\mathrm{ID}} (\fat{r}_{ij})
    = - \frac{\mu^2}{2 r^3_{ij}} .
\end{equation}
Clearly, for (\ref{eq:dipole_avg}) to be a good approximation to the true
inter-particle interaction, it is crucial that essentially all the particles
follow the field. An extensive analysis of this synchronization behavior of the
dipolar particles with the field in three dimensions can be found in
\cite{jaeger_klapp, jaeger_klapp_mhd}.

To systematically investigate the appearance of synchronization and clustering
we scanned a wide range of frequencies and field strengths. We consider a
system as clustered if the particles have on average more than $2.3$ neighbors
within a distance of $1.7 \sigma$ from their center. The latter value was used,
since it is slightly larger than the typical distance between neighboring
particles in a clustered system. This was found by looking at the first minimum
of respective pair correlation functions. The fact that we require $2.3$
neighbors on average ensures that two-dimensional aggregates are counted as
clusters while chainlike structures are disregarded.

\begin{figure}[ht]
    \centering
    \includegraphics[width=65mm]{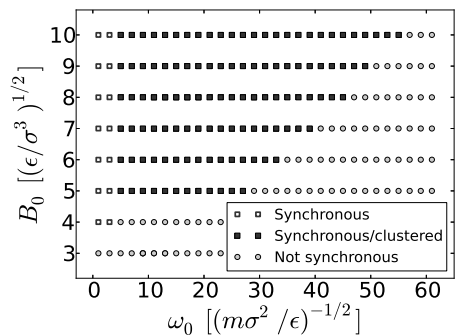}
    \caption{Synchronization behavior and cluster formation depending on
    frequency $\omega_0$ and strength $B_0$ of the field. The density and
    temperature of the system used were $\rho \sigma^2 = 0.2$ and $k_B T/
    \epsilon = 1.0$, respectively. $1225$ particles were used is these
    simulations.}
    \label{fig:sync}
\end{figure}

The results of this investigation of the space of the field parameters can be
seen in fig.~\ref{fig:sync}. Depicted are three distinct regions, denoted
``synchronous'', ``synchronous/clustered'', and ``not synchronous''. The first
region is comprised of systems in which the particles rotate synchronously with
the field but do not form clusters. Within this region, chains in the direction
of the field can be observed at low frequencies while spatial inhomogeneities
begin to appear at larger frequencies. As also becomes apparent here, a minimal
$B_0$ is required for the field to align the particles with itself.

In the second region, the synchronous rotation continues but is now accompanied
by the formation of two-dimensional clusters. This indicates that the effective
inter-particle potential becomes sufficiently isotropic within this region to
be reasonably described by the averaged dipolar potential
(\ref{eq:dipole_avg}).

As expected, the diagram in fig.~\ref{fig:sync} also depends on the friction
coefficients used in the BD simulations. In particular, we found the size of
the synchronous regime in the frequency domain to shrink in favor of the
``clustered'' regime if the translational friction coefficient $\xi_T$ is
increased. This results from reduced displacement of the particles after one
rotational period of the field thereby making eq.~(\ref{eq:dipole_avg}) valid
at much lower driving frequencies.

In the third region, we find neither synchronization nor cluster formation.
Clearly, the lack of synchronization is the direct cause of the breakdown of
cluster formation (cf.~eq.~(\ref{eq:dipole_avg})). The loss of synchronization
occurs, since the torques on the particles due to the external field become
unable to overcome the torques due to the rotational friction (similarly to
what is seen in three dimensions \cite{jaeger_klapp}).

Given the clustering scenario and our explanations so far, it is important to
ask to which extent the BD simulations can describe the dynamics of the real
colloidal system. The latter includes, by definition, a solvent. The rotating
external field, which constantly generates rotational motion of the particles,
will create flow fields that can result in considerable motion of the
particles. This might influence the cluster formation phenomenon due to the
following reason: The averaged potential (\ref{eq:dipole_avg}) is only valid as
an approximation to the true inter-particle interaction if the translational
motion of the particles during one rotational period of the field is small.

Hence, in order to find out whether cluster formation persists when
hydrodynamic interactions are present, we built a simulation that takes these
into account. To assess their influence on the cluster formation, it is
sufficient to consider only a single simulation box filled with particles.

\begin{figure}[ht]
    \centering
    \includegraphics[width=85mm]{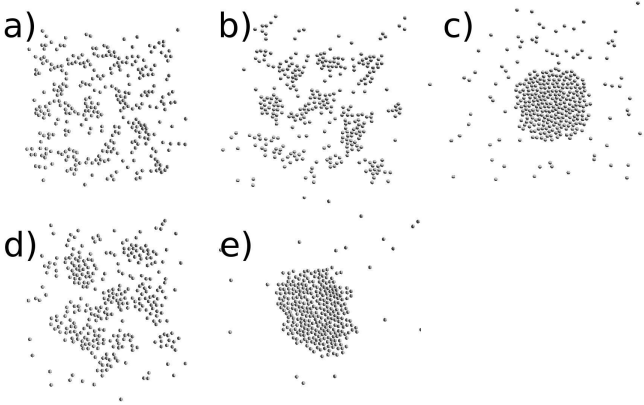}
    \caption{Snapshots of a system at different times without (top) and with
    (bottom) hydrodynamic interactions. The snapshots in the first column were
    taken at $t' = 2.5$, the ones in the second column at $t' = 10.8$, and the
    one in the last column at $t' = 138.5$ after the start of the simulation.
    Note that $t' = t D_0 / \sigma^2$}
    \label{fig:hydro_snaps}
\end{figure}

We considered a number of state points ($\rho \sigma^2 = 0.1, 0.2, 0.3$, $k_B
T/\epsilon = 1.0, 1.8$ at $\omega \sigma^2 / D_0 = 250$, $B_0
(\sigma^3/\epsilon)^{1/2} = 80)$ in our simulations with and without
hydrodynamic interactions included. We found that cluster formation occurs in
both these cases at all the considered densities and temperatures, with a
single cluster eventually forming in the simulation box. This can be seen in
fig.~\ref{fig:hydro_snaps}, where we show snapshots of the evolution of a
rotationally driven system. The snapshots in the top row show a system in which
hydrodynamic interactions are not taken into account, while hydrodynamic
interactions are present in the snapshots in the lower row. Another important
point that is illustrated by fig.~\ref{fig:hydro_snaps} is that the cluster
formation process is considerably accelerated by the hydrodynamic interactions.
At the intermediate time ($t' = 10.8$), only a single cluster remains in the
hydrodynamically interacting system (fig.~\ref{fig:hydro_snaps}e), while it
takes much longer for the not hydrodynamically interacting system to reach the
same state (cf.~fig.~\ref{fig:hydro_snaps}b,c).

Accelerated cluster formation is a direct consequence of both the
translation-translation and rotation-translation coupling. In particular the
former coupling accelerates the formation of clusters. This can be inferred
from selectively switching off the different hydrodynamic couplings.   Further,
note that if hydrodynamic interactions are present, the cluster rotates rapidly
around its center, which is not the case if these interactions are absent.
Here, the sole cause is the hydrodynamic rotation-translation coupling.

Thus, the hydrodynamic interactions affect the (rotational) cluster motion but
do not hinder the particles in the cluster formation process at all.

\section{Relation to condensation}
Given the clustering behavior of the non-equilibrium, yet fully synchronized,
field-driven system, we now ask to which extent the behavior of the system can
be understood by that of an {\it equilibrium} system interacting via the
effective interaction given by eq.~(\ref{eq:dipole_avg}) (as well as the
repulsive potential $U^{\mathrm{rep}}$ (cf.~eq.~(\ref{eq:interaction}))). More
precisely, our hypothesis is that the observed cluster formation stems from an
equilibrium, first-order phase transition between a vapor and a liquid phase.
To test this hypothesis, we have performed Wang-Landau-Monte-Carlo simulations
of a system interacting via $U^\mathrm{ID}$ (eq.~(\ref{eq:dipole_avg}))
\cite{footnote}.
\begin{figure}[ht]
    \centering
    \includegraphics[width=65mm]{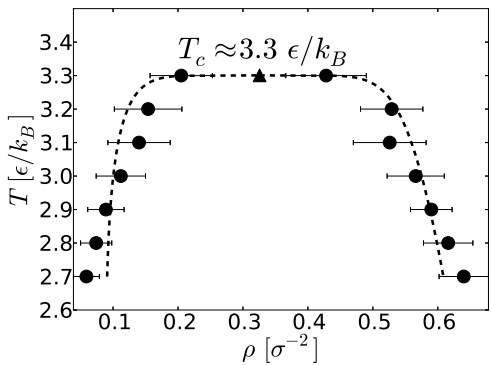}
    \caption{Phase diagram in the $T$-$\rho$ domain at $(\epsilon
    \sigma^3)^{1/2} \mu = 3$. $T_c$ denotes the critical temperature, the solid
    line corresponds to the binodal representing coexisting gas and liquid states.}
    \label{fig:phase_diagram}
\end{figure}

The key result we obtained from these simulations is the phase diagram
presented in fig.~\ref{fig:phase_diagram}. The presence of a binodal shows that
there is indeed a gas-liquid phase transition in the non-driven system.

Moreover, if we now compare this phase diagram with the thermodynamic state
point related to our non-equilibrium system (see e.g.~fig.~\ref{fig:snaps}b),
we find that the latter ($\rho \sigma^2 = 0.3$, $k_B T/\epsilon = 1$) lies well
within the two-phase region. This finding clearly suggests that the cluster
formation observed in the (fully synchronized) field-driven system is spinodal
decomposition, i.e., a dynamic phenomenon occurring during a first-order phase
transition. To check this conjecture in more detail, we looked at the time
evolution of the cluster sizes $\ell$. For phase separating systems it is well
established that $\ell$ exhibits power law behavior \cite{kapral2009},
i.e.~$\ell \propto t^\alpha$, with the corresponding exponents depending on the
growth stage. Such a behavior is also seen in Molecular dynamics (MD)
simulations. In particular, domains with growth proportional to $t^{1/2}$
\cite{koch1983, kabrede2006} and $t$ \cite{das2011} have been identified.

These power laws are universal in MD simulations but they do not necessarily
apply to BD simulations with their modified equations of motion. This was
shown, e.g., by Lodge and Heyes for the case of overdamped BD \cite{lodge1997}.
At the same time, however, the clusters in ref.~\cite{lodge1997} were still
found to grow with a power law. To check for the existence of cluster growth
with a power law in non-overdamped BD simulations, we first investigated a
``reference system'', whose equilibrium behavior is well studied. Specifically,
we considered a two-dimensional Lennard-Jones system at $\sigma^2 = 0.3$, $k_B
T/\epsilon = 0.45$ with the critical point being at $\rho \sigma^2 \approx
0.335$, $k_B T/\epsilon \approx 0.533$ \cite{Barker1981}. Investigating the
domain size, we did indeed find a power law dependence $\ell \propto t^\alpha$
with $\alpha \approx 0.30$. Note that the cluster size was obtained by
measuring the distance at which the pair correlation function assumes a value
of one for the first time if the radial bins are taken to be larger than the
particle diameters (cf.~\cite{desai1982}).

\begin{figure}[ht]
    \centering
    \includegraphics[width=80mm]{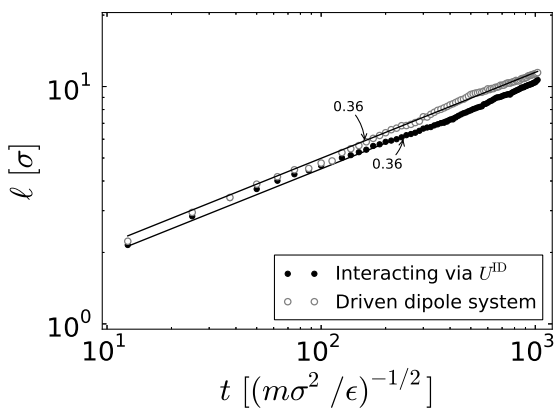}
    \caption{Cluster growths for a system interacting via the effective
    potential and a system that is driven by an external rotating field at
    density $\rho \sigma^2 = 0.3$ and temperature $k_B T/\epsilon = 1.5$ (state
    inside the coexistence region of fig.~\ref{fig:phase_diagram}). $4900$
    particles were used in the simulations.}
    \label{fig:cluster_size}
\end{figure}

Similarly, we checked the cluster growth for a driven dipolar system and a
system interacting via the potential $U^{\mathrm{ID}}$. In the simulations we
used a density and a temperature of $\rho \sigma^2 = 0.3$ and $k_B T/\epsilon =
1.5$, respectively, which put the systems well inside the coexistence region of
fig.~\ref{fig:phase_diagram}. The domain sizes over time that were extracted
from the simulations are shown in fig.~\ref{fig:cluster_size}. As can be seen,
the cluster sizes of these two systems are almost identical at any given time.
In particular, the characteristic domain sizes grow with a power law $\ell
\propto t^{\alpha}$ with $\alpha$ being equal to $0.36$ in both cases. We note
that this is almost identical to the Lifshitz-Slyozov growth law ($\ell \propto
t^{1/3}$) \cite{kapral2009}.

From these two results we conclude that the non-equilibrium system does indeed
undergo spinodal decomposition. First, the cluster growth proceeds with a power
law, which is typical within the spinodal region. Second, the growth behavior
remains unchanged even if the interactions between the driven dipoles are
replaced with the effective ones. This emphasizes the similarity between those
two systems and once more shows the validity of the phase diagram for the
driven system.

Note that hydrodynamic results are not included in fig.~\ref{fig:cluster_size}.
We only considered $324$ particles in these simulations and additionally did
not use periodic boundary conditions. This makes an accurate determination of
the cluster growth very difficult.

\section{Conclusions}
In this paper we have investigated the formation of two-dimensional aggregates
in monolayers of dipolar particles that are driven by rotating external
in-plane fields.

The first result of this paper is a non-equilibrium ``phase'' diagram, which
shows the regions of synchronization and cluster formation in the
$\omega-B_0$-domain. At high frequencies the synchronization of the particles
with the field breaks down, which results in a breakdown of cluster formation.
Similarly, we do not find cluster formation at low frequencies. Here, the
effective interaction between the particles is not well enough described by
$U^\mathrm{ID}$, since the particles move considerably during one rotational
period of the field. This changes in-between those frequencies, where the
particles rotate synchronously and sufficiently fast, which leads to the
formation of clusters.

Next, we investigated the stability of the clustering phenomenon when
hydrodynamic interactions are present. In our simulations, we found the
phenomenon to persist despite these interactions. In fact, the cluster
formation seems to proceed at even faster rates with these interactions
included. We attribute this to a combination of the hydrodynamic
translation-translation and rotation-translation coupling. The former seems to
accelerate the formation of clusters while the latter leads to the quick
formation of a single cluster in the center of the simulation box.

Test simulations indicate that these interactions have additional consequences:
Compared to a non-hydrodynamically interacting system, they seem to allow for
the formation of clusters at lower driving frequencies and therefore affect the
non-equilibrium ``phase'' diagram in fig.~\ref{fig:sync}. To study the precise
influence of the hydrodynamic interactions is outside of the scope of this
paper, but would be very illuminating in its own right.

We concluded our analysis with the main result of this letter: We established
the clustering phenomenon to be a by-product of a liquid-vapor phase
transition. This was done in two steps: We began by uncovering a phase
transition via Wang-Landau-MC simulations in a system interacting via the
effective potential $U^\mathrm{ID}$. Recall that this potential describes the
interactions between the particles very well in the driven system at
sufficiently high frequencies. In a next step we examined the domain growth of
the driven system within the binodal region of the phase diagram. As expected
within the spinodal region of a liquid-vapor phase transition, we found the
characteristic cluster size to grow with a power law. Additionally, it
essentially agrees with the domain growth of the non-driven system interacting
via the effective potential $U^\mathrm{ID}$. These facts lead us to conclude
that the clustering process corresponds to the pattern formation occurring
inside the binodal of a vapor-liquid phase transition.

Given these findings, it is interesting to compare them to recent experimental
results. Indeed, cluster formation in monolayers resulting from a rotating
external field has been observed multiple times \cite{weddemann2010,
wittbracht2011, snoswell2006, elsner, tierno2007}. In most of these
publications induced dipolar particles are brought to self-assemble into
two-dimensional aggregates. The only paper in which particles with a permanent
dipole moment were used (ref.~\cite{weddemann2010}) features a dipole-dipole
coupling strength ($\lambda_{\mathrm{DD}} = \mu^2 / k_B T \sigma^3$) that is
dominated by the dipole-field coupling strength ($\lambda_{\mathrm{DF}} = \mu
B_0/ k_B T$). There, the ratio $\lambda_{\mathrm{DF}}/\lambda_{\mathrm{DD}}$ is
about $6$, which is larger than the largest ratio that appears in
fig.~\ref{fig:sync}. Consequently, we expect the particles in
\cite{weddemann2010} to rotate synchronously with the field, resulting in the
effective interaction $U^\mathrm{ID}$ and the observed cluster formation.

The clusters found in the literature are typically hexagonally ordered. In our
simulations this becomes more and more true with increasing frequency and
strength of the field as well as with decreasing temperature. It should be
noted, however, that our BD simulations involve large Brownian kicks, which
result in deviations from the hexagonal structure. In the references
\cite{weddemann2010, wittbracht2011, snoswell2006, elsner}, micrometer-sized
particles were used, which makes these contributions much less significant
leading to more pronounced order.

Further, the driving frequencies used in these publications are considerably
smaller than the ones used here. This can once again be explained by the size
of the particles: Larger particles typically have smaller friction
coefficients, which, as test simulations show, result in cluster formation at
lower frequencies of the field.

The coarsening process investigated in this paper can universally be observed
in phase separating systems. Condensation transitions and spinodal demixing in
binary fluids or metallic alloys are popular examples of this. The process
reported in this letter is exceptional, however, in that the existence of a
liquid-vapour transition in ordinary dipolar system is still a hotly debated
topic and one of the big unresolved questions regarding these particles
\cite{ganzenmuller2009,tavares2011}. But as shown here, such a phase transition
can be induced via an external driving field. 

The system considered in this study is driven and, consequently, inherently in
a non-equilibrium state. The dynamic coarsening observed in spinodal
decomposition, on the other hand, is a process typically associated with
non-driven systems. It is, however, not unique to those. For instance, active
Brownian swimmers performing a ``run-and-tumble'' motion such as E.~Coli
bacteria, exhibit similar clustering behavior and demixing \cite{tailleur2008}.
With the ongoing and rising interest in dynamics and non-equilibrium processes
we expect an increasing amount of systems to be uncovered that are driven into
cluster or pattern formation with behaviors similar to the one described here.

\acknowledgements
We gratefully acknowledge financial support from the DFG within the RTG 1558
{\em Nonequilibrium Collective Dynamics in Condensed Matter and Biological
Systems} and IRTG 1524 {\em Self-Assembled Soft-Matter Nano-Structures at
Interfaces}.


\begin{thebibliography}{0}
\bibitem{footnote}
We calculated the energy contribution of (\ref{eq:dipole_avg}) via the well
know dipolar Ewald sum. This can be done, since $U^\mathrm{D}$ essentially
reduces to $U^\mathrm{ID}$, if the dipoles are oriented perpendicular to the
monolayer (cf.~ref.~\cite{smallenburg}).
\end{thebibliography}
\end{document}